\documentclass[12pt,nohyper,notoc]{JHEP}

\usepackage{amsmath,amssymb,cite,graphicx} 
\usepackage{epsf}
	
\newcommand{\beq}{\begin{eqnarray}}
\newcommand{\eeq}{\end{eqnarray}}

\font\zfont = cmss10 
\newcommand\ZZ{\hbox{\zfont Z\kern-.4emZ}}
\def\inbar{\vrule height1.5ex width.4pt depth0pt}
\def\IC{\relax\hbox{\kern.25em$\inbar\kern-.3em{\rm C}$}}

\newcommand{\EQ}[1]{\begin{equation} #1 \end{equation}}

\title{The Effective Lagrangian in the Randall-Sundrum Model and Electroweak
Physics}

\author{Csaba Cs\'aki$^{a}$, 
Joshua Erlich$^b$, and John Terning$^b$\\
$^a$Newman Laboratory of Nuclear Studies, Cornell University, Ithaca, NY 14853,
USA
\\
$^b$Theory Division T-8, Los Alamos National Laboratory, Los Alamos,
NM 87545, USA
\\
Email: {\tt csaki@mail.lns.cornell.edu, erlich@lanl.gov, terning@lanl.gov}}

\abstract{
We consider the two-brane Randall-Sundrum (RS) model with bulk gauge fields.
We carefully match the bulk theory to a 4D low-energy effective Lagrangian.
In addition to the four-fermion operators induced by KK exchange we find
that large negative $S$ and $T$ parameters are induced in the effective
theory. This is a tree-level effect and is a consequence of the shapes of the
$W$ and $Z$ wave functions in the bulk. Such effects are generic
in extra dimensional theories where the 
standard model (SM) gauge bosons have non-uniform wave functions along the
extra dimension. The corrections to
precision electroweak observables in the RS model are mostly
dominated by $S$. We fit the parameters of the RS model 
to the experimental data and find somewhat stronger bounds than 
previously obtained; however, the standard model bound on the 
Higgs mass from precision measurements can only be slightly relaxed 
in this theory. 
}

\preprint{{\tt hep-ph/0203034}\\
CLNS 02/1778}

\begin{document}

\section{Introduction}

Theories with extra dimensions might explain some of the outstanding problems
of particle physics \cite{ADD,RS,RS2}. In particular some of these models
could shed light on 
why gravity is so much weaker than the other three forces. One of the prominent
proposals of this sort is the Randall-Sundrum (RS) 
model~\cite{RS,RS2}, where the strong warping 
of the extra dimensions introduces an exponential hierarchy between the 
Planck and the weak scales. There are several variants of this model,
depending on whether the extra dimension is finite (RS1) or infinite (RS2),
and whether or not the gauge fields are in the bulk. Each of these
models can be interesting for slightly different motivations. Here we will
concentrate on the case where the extra dimension is finite (so that it 
solves the hierarchy problem), and where the gauge fields are in the bulk.
This model could possibly yield unification of gauge 
couplings~\cite{RSchwartz},
and also may have a simple physical origin~\cite{holography,Riccardo} 
via the AdS/CFT
correspondence~\cite{Juan}. 
The holographic dual of this theory should be a broken
conformal field theory, which  becomes strongly interacting at low
energies and spontaneously breaks the weakly gauged $SU(2)\times U(1)$ 
electroweak symmetries. Thus this holographic dual of the RS model
with the gauge bosons in the bulk is in essence a technicolor-like theory, 
where the
broken CFT replaces the technicolor group~\cite{holography}, and the 
KK modes of the gauge fields and gravitons would be interpreted 
as bound states of the CFT resulting in the technimesons, analogously
to the glueball states appearing in the case of ordinary 
AdS/CFT~\cite{glueball}. In QCD-like technicolor theories 
the new strong interactions introduced to solve the hierarchy problem
always generate large contributions to the electroweak precision 
observables~\cite{STC}, in particular there are large contributions to
the $S$-parameter.
However the phenomenological importance of (non-QCD-like) 
approximate conformal
symmetry has long been emphasized in the technicolor 
literature \cite{walking}, where
the slowly running gauge coupling is refered to as ``walking''.
The difficulty of estimating the value of $S$ in these walking
theories is also well known \cite{Lane} since it involves non-perturbative,
non-supersymmetric gauge dynamics near a non-trivial fixed point.
Therefore it is interesting to find out whether there is
a non-vanishing $S$-parameter in the RS model since it provides
us with the first approximately conformal (``walking'') model of
electroweak symmetry breaking where such a calculation can be performed.
However, in the 5D gravity theory (that is the RS model) the $S$ parameter
should not be the effect of quantum loops, but rather a purely tree-level
effect. The purpose of this paper is to carefully match the RS model
to an effective 4D description and find the value of $S$ in
the effective Lagrangian
describing electroweak physics in this model. 
Indeed we find that the wave functions of the $W$ and $Z$ bosons are 
distorted due to the Higgs expectation values on the TeV brane, resulting
in different wave function and mass renormalizations of the $W$ and $Z$.
The physical consequence of this effect is  non-vanishing $S$ 
and $T$ parameters, which we calculate. Our method of finding the  
low-energy effective 4D theory is general, and we expect that similar 
effects will appear in any extra dimensional theory where the SM gauge bosons 
have non-uniform wavefunctions. In addition to these parameters
the well-known effect of the four-fermion operators generated by the 
exchange of Kaluza-Klein gauge bosons has to be included. The 
coefficient of these four-fermion operators has been called $V$ in 
\cite{RizzoWells,Hooman}.
We use a global fit to the most recent
precision electroweak data to place a bound
on the size of the extra dimension in the RS model, and find bounds that are 
somewhat stronger than those previously obtained.

The paper is organized as follows: in Section 2 we review the results
on gauge propagators and wave functions in the RS model that will
be necessary to calculate the effective Lagrangian. In Section 3, we
match the higher dimensional theory to an effective 4D Lagrangian, and
evaluate the $S$ and $T$ parameters. In Section 4 we first calculate the $V$
parameter and then use these results
for constraining the parameters of the RS model 
via a global fit to the electroweak precision measurements. We conclude 
in Section 5, while Appendix A contains the detailed expressions of the
electroweak observables in terms of $S,T$ and $V$ and the SM input and
experimental values used for our fit.

\section{The Gauge Propagator and Wavefunctions in the RS Model}

In this section we review the results on gauge propagators and wave functions
in the RS model [4,12-17]
that will be necessary for us to calculate the effective 
low-energy theory.

The 5D metric of the RS model can be written in the form
\EQ{
ds^2=\left(\frac{R}{z}\right)^2\left(-dz^2+\eta_{\mu\nu}dx^\mu\,dx^\nu \right) ,
\label{nmetric}
}
for $R < z <R'$. Here $R$ represents the radius of curvature of the AdS space.
There is a Planck brane at $z=R$ and a TeV brane at $z=R'$ which cutoff the space
with $\ZZ_2$ orbifold boundary conditions.

The 5D action for the bulk $SU(2)\times U(1)$ gauge bosons is given by
\beq
S_{5D} &=& \int d^4x \int_R^{R'} dz 
\sqrt{-G} \left[-\frac{1}{4 g_5^2} G^{MP} G^{NQ}
  W^a_{M N }W^a_{P Q}
-\frac{1}{4 g_5^{\prime\,2}} G^{MP} G^{NQ} B_{M N } B_{P Q} 
\right.\nonumber \\
&&\left.+\frac{v^2}{8}\frac{\delta(z-R')}{\sqrt{G_{55}}} 
G^{MP} \big( W^1_M W^{1}_P+W^2_M W^{2}_P +(W_M^3-B_M)(W_P^3-B_P)\big)
\right] ,
\eeq
where the $\delta$-function mass terms arise from a localized
Higgs expectation
value $\langle H\rangle =v/2$.  Since $\delta$-functions on boundaries
require special care, we will take the definition of this term to 
be the limit of having the Higgs localized at a point which approaches the
brane. This amounts to a factor of two difference in the definition of
$v^2$ from taking the Higgs directly on the brane, but has the advantage of
making comparison with the calculation via the
AdS/CFT correspondence simpler~\cite{cetwip}.
 
We define the weak mixing angle through $s$,
which  represents the bare value of the $\sin \theta_W$:
\beq
s= \frac{g_5'}{\sqrt{g_5^2+g_5^{\prime\,2}}},\ \ \ \ 
c=\frac{g_5}{\sqrt{g_5^2+g_5^{\prime\,2}}} ~.
\eeq
We can diagonalize the action by
performing a field redefinition:
\beq
W^3_\mu = c^2 Z_\mu +  A_\mu\,, \ \ \ \
B_\mu = -s^2 Z_\mu +  A_\mu ~.
\eeq
The reason behind this unusual form for the field redefintions is that
none of the fields $W^3,B,Z$ or $A$ are canonically normalized, but it
is equivalent to the standard redefinition in the canonical basis.
In our new basis we obtain the Lagrangian:
\beq
S_{5D} &=&  \int d^4x \int_R^{R'} dz 
\frac{R}{z} \left[-\frac{1}{2 g_5^2} 
  W^+_{M N }W^{-M N}
-\frac{1}{4}\left(\frac{1}{g_5^2}+\frac{1}{ g_5^{\prime\,2}}\right)
  F_{M N } F^{M N} \right.\nonumber \\
&&\left.-\frac{1}{4 (g_5^2+g_5^{\prime\,2})}  Z_{M N } Z^{M N}
+\frac{v^2}{4} \delta(z-R') \frac{R}{z}
 W^+_M W^{-M} +\frac{v^2}{8} \delta(z-R')\frac{R}{z}  Z_M Z^{M} \right]
\eeq
In the $R_\xi$ gauge where $W_5=Z_5=A_5=0$,
the propagator for the bulk $W$ gauge boson is given by~\cite{RSchwartz},
\beq
\Delta_W^{\mu \nu}&=&(\eta^{\mu \nu}- q^\mu q^\nu/q^2)\,\Delta_W(q,z,z')+
\Delta_{W,\xi}\,
q^\mu q^\nu/q^2,
\eeq
where \beq
\Delta_{W,\xi}=\Delta_W(q/\sqrt{\xi},z,z'), \eeq
and $\Delta_W$ satisfies
\beq
 g_5^2 \frac{z}{R} \delta(z-z')&=&(\partial_z^2 - \frac{1}{z} \partial_z  +q^2 -\frac{1}{4}
v^2 g_5^2  \delta(z-R') \frac{R}{R'}) 
\Delta_W(q,z,z')
\label{greeneq}
\eeq
with boundary conditions 
\beq
\partial_z \Delta_W|_{z=R} = 0 ~, \ \ \ \ 
\partial_z \Delta_W|_{z=R'} = 
-\frac{1}{4}  g_5^2 v^2  \frac{R}{R'} \Delta_W ~.
\label{bc}
\eeq

$W^5$ also propagates in a generic $R_\xi$ gauge, but the coupling to
fermions is pseudoscalar, and vanishes at zero fermion mass.  In addition, 
because $W^5$ is fixed to be odd under the $\ZZ_2$ (consider the $\ZZ_2$ 
behavior of $W_{5\mu}$), it vanishes on the TeV brane and hence does not
couple to matter on the TeV brane.  For our purposes we 
will therefore be able to neglect $W^5$.

For $R\le (z,z')\le R'$ the Green's function can be constructed by patching
together the
solutions of the corresponding homogeneous
equation with $z<z'$ and $z>z'$, which we refer to as $\Delta_{W<}$ and
$\Delta_{W>}$ respectively:
\beq
 \Delta_W = \theta(z-z') \Delta_{W>} + \theta(z'-z) \Delta_{W<}
\eeq
Plugging the patched solution into Eq.~(\ref{greeneq}) for $z'\ne R$, 
$z'\ne R'$
yields:
\beq
&&\Delta_{W<}|_{z=z'} = \Delta_{W>}|_{z=z'}
\label{patch1} \nonumber \\
&&\partial_z(\Delta_{W>}-\Delta_{W<})|_{z=z'}= g_5^2  \frac{z'}{R}
\label{patch2}
\eeq
Setting $z'=R'$ in Eq.~(\ref{patch2}) and combining
with Eq.~(\ref{bc})  yields the IR ($z=R'$) 
boundary condition for the propagator with
a source on 
the TeV brane:
\beq
\partial_z \Delta_{W<}|_{z=z'=R'}=-\frac{R'}{R} g_5^2 
-\frac{1}{4} g_5^2 v^2 \frac{R}{R'} \Delta_{W<}|_{z=z'=R'}
\label{norm}
\eeq
The solution has the form
\beq
\Delta_W(q,R',z)=\Delta_{W<}|_{z'=R'}= \frac{z}{R}\left(\alpha_W  J_1(q z) 
+ \beta_W Y_1(q z)\right)
\eeq
where the coefficients are given by
\begin{equation}
\alpha_W= \frac{4 g_5^2 R' Y_0(q R)}{D(q)}, \ \ \ \ 
\beta_W = \frac{-4 g_5^2 R' J_0(q R)}{D(q)} 
\end{equation}
and the denominator is 
\begin{equation}
D(q)=J_0(q R)\left(4 q R' Y_0(q R')+g_5^2 v^2 R Y_1(q R')\right)
-Y_0(q R)\left(4 q R' J_0(q R')+g_5^2 v^2 R J_1(q R')\right).
\end{equation}
Setting $z=R'$ gives the propagator on the TeV brane.
The coefficients $\alpha_W$ and $\beta_W$ 
have the same denominators, and the roots of these
denominators determine the 4D poles of the propagator (when the numerators
do not vanish concurrently). The $n^{th}$ pole
corresponds to
the $n^{th}$ $W$ eigenmode
, $W^{(n)}$, with mass $M_W^{(n)}$.
We will label the lowest mode by $n=0$.
It is this lowest mode that we would like to identify with the 
observed $W$ gauge
boson, so we will write $M_W^{(0)}=M_W$. 
For $v R \ll 1$  
this is an ``almost zero mode'' and we can find the pole analytically
by expanding the Bessel functions in $q R <  q R' \ll 1$.
To leading order in the coupling the pole is at:
\beq
M_W^2 \approx \frac{g_5^2} {R \log(R'/R)}\frac{R^2 v^2}{4 R^{\prime\,2}}
\label{MW}
\eeq
The bulk $Z$ propagator is obtained from $\Delta_W$ by taking 
$g_5^2\rightarrow g_5^2 + g_5^{\prime 2}$, which gives:
\beq
M_Z^2 \approx \frac{g_5^2+g_5^{\prime\,2}} {R \log(R'/R)}\frac{R^2 v^2}{4 R^{\prime\,2}}
\label{MZ}
\eeq

Similarly we can find the wavefunction, $\psi^{(n)}_W$,
of the $n^{th}$ eigenmode.   The wavefunctions are given by:
\beq
\psi_W^{(n)}(z) = \frac{z}{R'} \frac{J_1(M_W^{(n)}z) Y_0(M_W^{(n)}R)-
Y_1(M_W^{(n)}z) J_0(M_W^{(n)}R)}{J_1(M_W^{(n)}R') Y_0(M_W^{(n)}R)-
Y_1(M_W^{(n)}R') J_0(M_W^{(n)}R)}
\eeq
where we have normalized the wavefunction by $\psi^{(n)}(R')=1$.

For $v R \ll 1$
the $n=0$ mode is an ``almost zero mode'' and we can find a simple expression
for the wavefunction
by expanding the Bessel functions in $M_W R \ll 1$, $M_W R' \ll 1$. To order
$M_W^2$ we obtain:
\beq
\psi_W^{(0)}(z) &\approx& 1+ \frac{M_W^2}{4} \left[ z^2-R^{\prime\,2}-2 z^2 \log(z/R)
+2 R^{\prime\,2} \log(R'/R) \right].
\eeq
The wavefunction for the $Z$ can be obtained by 
$M_W\rightarrow M_Z$,
while the photon remains massless and its wavefunction is simply 
$\psi_\gamma^{(0)}(z)=1$.

\section{The 4D Effective Lagrangian of the RS Model}

Given the wavefunctions we can find the 4D effective action by integrating
the 5D action over $z$. 
We want to match the RS calculation onto an effective 4D theory.
After integrating out the Higgs, the most general Lagrangian for the 
electroweak gauge bosons (with operators of dimension 4 or less) can
be written as~\cite{STC,Holdom,Georgi}:
\beq
{\cal L} &=& -\frac{1}{2 g^2}Z_W  W^+_{\mu \nu}W^{- \mu \nu}
-\frac{1}{4 (g^2+g^{\prime\,2})}Z_Z  Z_{\mu \nu} Z^{ \mu \nu}
-\frac{1}{4 e^2}Z_\gamma  F_{\mu \nu} F^{ \mu \nu}
+\frac{sc}{2 e^2}\Pi^\prime_{\gamma Z}  F_{\mu \nu}Z^{\mu \nu} \nonumber\\
&&+\left(\frac{f^2}{4}+\frac{1}{ g^2}\Pi_{WW}(0)\right) W^+_\mu W^{-\mu}
+\frac{1}{2}\left(\frac{f^2}{4}+\frac{1}{ (g^2+g^{\prime\,2})}\Pi_{ZZ}(0)\right) 
Z_\mu Z^\mu
\eeq
where $Z_\gamma \equiv 1-\Pi^\prime_{\gamma \gamma}$, 
$Z_W \equiv 1-\Pi^\prime_{W W}$, 
$Z_Z \equiv 1-\Pi^\prime_{Z Z}$, 
$\Pi^\prime_{\gamma Z}$, $\Pi_{WW}(0)$, 
and $\Pi_{ZZ}(0)$
incorporate the effects of new (oblique) physics beyond the standard model.
Although we are only doing a tree-level matching
calculation we have adopted the
standard notation for vacuum polarizations to represent the wavefunction
renormalizations that arise from classical 5D physics.
With our conventions $f \approx 246$ GeV.

 Using the 5D wavefunctions (to second order in masses) we can easily
calculate the coefficients of the kinetic and mass terms:
\beq
\frac{1}{e^2}Z_\gamma &\equiv&
\left(\frac{1}{g_5^2}+ \frac{1}{g_5^{\prime\,2}}\right) 
\int_R^{R'} |\psi_\gamma^{(0)}(z)|^2 \frac{R dz}{z} =
 \left(\frac{1}{g_5^2}+ \frac{1}{g_5^{\prime\,2}}\right) R \log(R'/R)
\nonumber \\
\frac{1}{g^2}Z_W &\equiv& \frac{1}{g_5^2} 
\int_R^{R'} |\psi_W^{(0)}(z)|^2 \frac{R dz}{z}
=\frac{1}{g_5^2} R \log(R'/R)-\Pi_{11}^\prime
\nonumber \\
\frac{1}{g^2+g^{\prime\,2}} Z_Z &\equiv& 
\frac{1}{g_5^2+g_5^{\prime\,2}} \int_R^{R'} |\psi_Z^{(0)}(z)|^2 \frac{R dz}{z}
=\frac{1}{g_5^2+g_5^{\prime\,2}} R \log(R'/R)
-\Pi_{33}^\prime
\nonumber \\
\Pi^\prime_{\gamma Z}&=& 0 \nonumber \\
\frac{f^2}{4} + \frac{1}{ g^2} \Pi_{WW}(0)&\equiv& 
\frac{R^2}{4R^{\prime\,2}} v^2+ \frac{1}{g_5^2} 
\int_R^{R'} |\partial_z\psi_W^{(0)}(z)|^2 \frac{R dz}{z}\nonumber
= \frac{R^2}{4R^{\prime\,2}} v^2 + \Pi_{11}(0)  \\
\frac{f^2}{4} + \frac{1}{ (g^2+g^{\prime\,2})} \Pi_{ZZ}(0)&\equiv& 
\frac{R^2}{4R^{\prime\,2}} v^2 + 
\frac{1}{g_5^2+g_5^{\prime\,2}} \int_R^{R'} |
\partial_z\psi_Z^{(0)}(z)|^2 \frac{R dz}{z}=
\frac{R^2}{4R^{\prime\,2}} v^2 + 
\Pi_{33}(0)
\eeq
where
\beq
\Pi_{11}^\prime&=&\Pi_{33}^\prime=
- \frac{R^2 v^2}{8R^{\prime\,2}}\left(2 R^{\prime\,2} \log(R'/R) 
-2R^{\prime\,2} +  \frac{R^{\prime\,2} -R^2}{   \log(R'/R)} \right)\nonumber \\
\Pi_{11}(0)&=&\frac{g_5^2 R^3 v^4}{64 R^{\prime\,4}}\left(2 R^{\prime\,2} 
-\frac{2R^{\prime\,2}}{\log(R'/R)} +\frac{R^{\prime\,2} -R^2}{\log(R'/R)^2} \right)
=-\frac{M_W^2}{2} 
\Pi_{11}^\prime\nonumber \\
\Pi_{33}(0)&=&\frac{(g_5^2+g_5^{\prime\,2}) R^3 v^4}{64 R^{\prime\,4}}
\left(2 R^{\prime\,2} 
-\frac{2R^{\prime\,2}}{\log(R'/R)} +\frac{R^{\prime\,2} -R^2}{\log(R'/R)^2} \right)
=-\frac{M_Z^2}{2} 
\Pi_{11}^\prime~.
\eeq
Here we have used the leading order results for $M_W$ and $M_Z$, 
Eqs.~(\ref{MW}). The corrections to the wave function 
renormalization $\Pi'_{11}$ and $\Pi'_{33}$ arise from integrating the
$W$ and $Z$ wave functions, while the contributions to the mass 
renormalizations $\Pi_{11}$ and $\Pi_{33}$ appear from the 5D kinetic terms
of $W$ and $Z$, where a $z$ derivative acts on the wave functions.

Even though there are no $1/(16 \pi^2)$ loop suppression factors,
for $v R \ll 1$, $\Pi_{11}$ and $\Pi_{33}$ can be treated as small perturbations
of the leading terms. Thus a simple convention is to identify the 4D bare
gauge couplings with the leading terms and the $Z_i-1$  with
the subleading $\Pi^\prime$ terms.  This is a convenient choice because
we want to separate out the new 5D physics from 
radiative corrections by loops of standard model particles. It also 
ensures that there are no additional corrections to the couplings of 
the $W$ and $Z$ to quarks and leptons. In other words, with this
convention the mixing angles
determined by diagonalizing the 5D action are identical to the 
bare 4D mixing angles:
\beq
s= \frac{g'}{\sqrt{g^2+g^{\prime\,2}}},\ \ \ \ 
c= \frac{g}{\sqrt{g^2+g^{\prime\,2}}} ~.
\eeq
It is these mixing angles that appear in the quark and lepton gauge couplings.
Other conventions are possible, but physical observables
are independent of convention. Thus in our convention
\beq
\frac{1}{e^2} &\equiv&
 \left(\frac{1}{g_5^2}+ \frac{1}{g_5^{\prime\,2}}\right) R \log(R'/R) \nonumber \\
\frac{1}{g^2} &\equiv& \frac{1}{g_5^2} R \log(R'/R)\nonumber \\
\frac{1}{g^2+g^{\prime\,2}}  &\equiv& \frac{1}{g_5^2+g_5^{\prime\,2}} R \log(R'/R).
\eeq
Using this identification of the 4D couplings we then obtain for the 
other parameters of the effective Lagrangian:
\beq
 Z_\gamma&=&1~, \ \ \ \ \ \ \ \ \  
Z_W = 1-g^2 \Pi_{11}^\prime~, \ \ \ \ \ \ \ \ 
Z_Z = 1-(g^2+g^{\prime\,2}) \Pi_{33}^\prime~, \nonumber \\
f^2&=& \frac{R^2}{R^{\prime\,2}} v^2~, \ \ \ 
\Pi_{WW}(0)=g^2 \Pi_{11}(0)~, \ \ \ 
\Pi_{ZZ}(0)=(g^2+g^{\prime\,2}) \Pi_{33}(0)~.
\eeq

Note that since the photon is massless it receives no 5D renormalization,
so $\Pi_{\gamma \gamma}=e^2\Pi_{QQ}=0$. Furthermore since this is a tree-level calculation no new $Z-\gamma$ mixing
can be induced so $\Pi_{3Q}=0$.
We can now use the standard definitions\cite{STC,Sformulas} for the oblique
parameters:
\beq
S&\equiv&16 \pi(\Pi_{33}^\prime - \Pi_{3Q}^\prime)\nonumber \\
T&\equiv&\frac{4 \pi}{s^2 c^2 M_Z^2} (\Pi_{11}(0) - \Pi_{33}(0))\nonumber\\
U&\equiv&16 \pi(\Pi_{11}^\prime - \Pi_{33}^\prime)
\eeq
Plugging in our results yields:
\beq
S&\approx& - 4 \pi  f^2 R^{\prime\,2}  \log(R'/R)\nonumber \\
T&\approx& -\frac{\pi}{2 c^2}  f^2 R^{\prime\,2}  \log(R'/R)~, \nonumber \\
U&=&0~,
\eeq
where we have dropped terms which are suppressed by powers of $\log(R'/R)$.
Note that both $S$ and $T$ are negative and large (i.e.
$\log$ enhanced relative
to a naive dimensional analysis estimate).

We can check these results by examining the poles of the propagators
directly by expanding the denominators for $q R \ll 1$.
At leading order the $W$ pole is determined by
\beq
0= -\frac{1}{4} \frac{R^2}{R^{\prime\,2}} v^2 + q^2
\frac{R \log(R'/R)}{g_5^2}\,.
\eeq
At next to leading order we must keep terms that are suppressed by
$q^2 R^{\prime\,2}$ relative to the leading terms.
The pole is then determined by
\beq
0= -\frac{1}{4} \frac{R^2}{R^{\prime\,2}} v^2 + q^2
\frac{R \log(R'/R)}{g_5^2} + A q^2 +B q^4 
\eeq
where
\beq
A&=& \frac{R^2 v^2}{16 R^{\prime\,2}}\left(2 R^{\prime\,2} 
\log(R'/R)+R^2 -R^{\prime\,2}\right)~, \nonumber \\
B&=&- \frac{R }{4 g_5^2 }
\left(( R^{\prime\,2}+R^2) \log(R'/R)+R^2-R^{\prime\,2}\right)\,.
\eeq
Thus to sub-leading order the pole is at:
\beq
M_W^2 &\simeq& \frac{g_5^2} {R \log(R'/R)}\frac{R^2 v^2}{4 R^{\prime\,2}}
- \frac{g_5^4 R^2 v^4}{64  R^{\prime\,4} \log(R'/R)^3 }\left(2 R^{\prime\,2} \log(R'/R)^2
-2 R^{\prime\,2} \log(R'/R) +R^{\prime\,2}\right)\nonumber\\
&\simeq&\frac{g^2 f^2}4
- \frac{g^4  f^4}{64 }\left(2 R^{\prime\,2} \log(R'/R)^2
-2 R^{\prime\,2} \log(R'/R) +R^{\prime\,2}\right),
\label{MWsubleading}
\eeq
which agrees at this order
with the effective Lagrangian calculation 
\beq
M_W^2 \simeq g^2 \left(\frac{f^2}{4}
+ \Pi_{11}(0) \right)\left(1+g^2 \Pi_{11}^\prime \right)
\simeq g^2 \frac{f^2}{4}\left(1+\frac{g^2}{2} \Pi_{11}^\prime \right)~.
\eeq

\section{The Comparison of RS to Data}
In addition to the oblique corrections we have described, once quarks
and leptons are
included in the theory with couplings to the bulk gauge bosons
they will have additional four-fermion interactions beyond those in the
standard model
due to the exchange of the gauge boson resonances. The effect of these
corrections has been parameterized in ref.~\cite{RizzoWells,Hooman}  
by a correction
to $G_F$ denoted by $V$.  Recall that the 
contribution  to $G_F$ from $W$ exchange is~\cite{Sformulas}:
\beq
4 \sqrt{2} G_{F,W}= \frac{1}{\frac{f^2}{4} +\Pi_{11}(0)}~.
\label{treeGF}
\eeq
To include the effect of resonances we can write the bulk
$W$ propagator as a sum over poles:
\beq
\Delta_W(q,R',R')= g_5^2 \sum_{n=0}^\infty \frac{\psi_W^{(n)}(R')^2}{N_n(q^2-M_W^{(n)2})}
\eeq
where $N_n$ is determined by 
\beq
N_n= \int_R^{R'} dz |\psi_W^{(n)}(z)|^2~,
\eeq
and hence 
\beq
N_0 = Z_W R \log(R'/R)~.
\eeq
Since we chose $\psi_W^{(n)}(R') =1$, we then have 
\beq
 \Delta_W(q,R',R')
= \frac{1}{Z_W}\frac{g^2}{q^2-M_W^2}+ 
g^2 \sum_{n=1}^\infty \frac{N_0 \psi_W^{(n)}(R')^2}{Z_W N_n(q^2-M_W^{(n)2})}.
\label{eq:Gfermi}\eeq
At zero momentum the first term on the right hand side is just 
$-4 \sqrt{2} G_{F,W}$, and the remaining terms are the additional corrections
not coming from the $W$ pole.
If we write the correction to $G_F$ as $G_F=G_{F,W}(1+V)$ then we have
\beq
V&=&-\left( \Delta_W(q=0,R',R')
+\frac{1}{\frac{f^2}{4} +\Pi_{11}(0)}\right)
\left(\frac{f^2}{4} +\Pi_{11}(0)\right) \nonumber \\
&\simeq& \frac{g_5^2 R v^2}{16 R^{\prime\,2}}\left(2 R^{\prime\,2} 
-\frac{2R^{\prime\,2}}{\log(R'/R)} +\frac{R^{\prime\,2} -R^2}{\log(R'/R)^2} \right)
\simeq \frac{g^2}{8}f^2 R^{\prime\,2} \log(R'/R)~,
\eeq
where in the last line we have again dropped terms suppressed by powers of
$\log(R'/R)$.

Thus we see that there are three types of corrections to precision  
electroweak observables in RS models: $S$, $T$, and $V$. Note that
some of these corrections have also been considered in Refs. 
\cite{HuberShafi,Huber}.
To relate our parameters to observables we use the 
standard definition of
$\sin \theta_0$ from the $Z$ pole, 
\beq
\label{s2}
\sin^2  \theta_0 \cos^2  \theta_0 &=& \frac{\pi \alpha(M_Z^2)}{\sqrt{2} G_F M_Z^2}~,\\
\sin^2 \theta_0&=&0.23105 \pm 0.00008
\eeq
where~\cite{ErlerLang} 
$\alpha(M_Z^2)^{-1}=128.92\pm 0.03$ is the running SM fine-structure
constant
at $M_Z$.
We can relate this measured value with the bare value 
in this class of models, 
\beq
\sin^2  \theta_0 = s^2 +\frac{s^2 c^2}{c^2-s^2} \left( - 
\frac{\alpha}{4 s^2 c^2} S +\alpha T -V\right),
\label{rens2}
\eeq
which is obtained by considering all corrections to (\ref{s2}) in the 
usual way (see~\cite{Sformulas}).
Also, in the RS model we have the simple result that with only
the tree-level 5D renormalizations the
running couplings defined by Kennedy and Lynn~\cite{Lynn}
which appear in $Z$-pole 
asymmetries are the same as
the bare couplings:
\beq
s^2_*(q^2)=s^2, \ \ \ 
e^2_*(q^2)=e^2 ~.
\eeq

In addition to the contribution from  $T$, 
there are further corrections to the
low-energy ratio of charged- to neutral-current interactions coming
from resonance exchange.
We will absorb this effect into the parameter $\rho_*$:
\beq
\rho_*=
\frac{\frac{f^2}{4} +\Pi_{11}(0) }{\frac{f^2}{4} +\Pi_{33}(0)} 
\left(\frac{1+V/c^2}{1+V}\right)
\approx 1 + \alpha T+ \frac{s^2}{c^2} V ~.
\label{rho*}
\eeq
Curiously in the RS model we find that the contributions from $T$ and $V$
cancel, and $\rho_*=1$.  We will however present the general results
for precision observables
in  Appendix A without assuming a relation between $T$ and $V$, so
that our results can be used for more general models.
With Eqs.~(\ref{rens2}) and (\ref{rho*}) it is straightforward to calculate
the corrections in a general model
to precision electroweak observables in terms of $S$, $T$, and
$V$. The expressions for the various observables 
together with the SM predictions and experimental results are given 
in Appendix A. 

The result of a global fit to the 23 observables 
listed in Table \ref{table} is that for 
$M_{\rm Higgs}=115$ GeV 
\beq
R' (\log(R'/R))^{1/2} < 0.50\, {\rm TeV}^{-1}
\eeq
at the 95 \% confidence level. Taking $\log(R'/R) = 32$ (as is often
done to naturally explain the hierarchy between the Planck and weak
scales~\cite{RS,RS2,RSchwartz}) we have:
\beq
1/R'  > 11 \,{\rm TeV}~.
\label{bound1}
\eeq
For a value of $R'$ which saturates this bound we have
\beq
S&=& -0.19 \nonumber\\
T&=& -0.03 \nonumber\\
V&=& 0.00082 
\eeq
Since the $T$ and $V$ contributions to $\rho_*$ cancel and 
for these values the contribution of
$S$ to $(s^2-\sin^2 \theta_0)$ is about 8.6 times larger than that of $T$ and
2.6 times larger than that of $V$, it is the $S$ parameter constraint
that dominantly determines the bound on $R'$.

It is interesting to note that for
the $Z$-pole observables, which do not depend on $\rho_*$, one can
absorb the contribution from $V$ into an effective $T$:
\beq
T_{\rm eff}= T- \frac{V}{\alpha}~.
\eeq
Thus in the RS model, $T_{\rm eff}$ is even more negative than $T$. 
One can then use the bounds on $S$ and $T$ to 
estimate the bounds on $R'$, which yields results similar to 
(\ref{bound1}).

In the RS model there is also a light radion that contributes to precision
electroweak observables at the loop level 
(together with loops of the KK gravitons and gauge bosons).  
These contributions have
been calculated separately~\cite{CGK} and are 
small unless an extra Higgs-radion coupling~\cite{GRW} 
is introduced.  With this  additional coupling, the radion corrections tend to
make $S$ more negative and hence only tighten the bound on $R'$.

\begin{figure}[ht]
\centerline{\epsfxsize=10cm\epsfbox{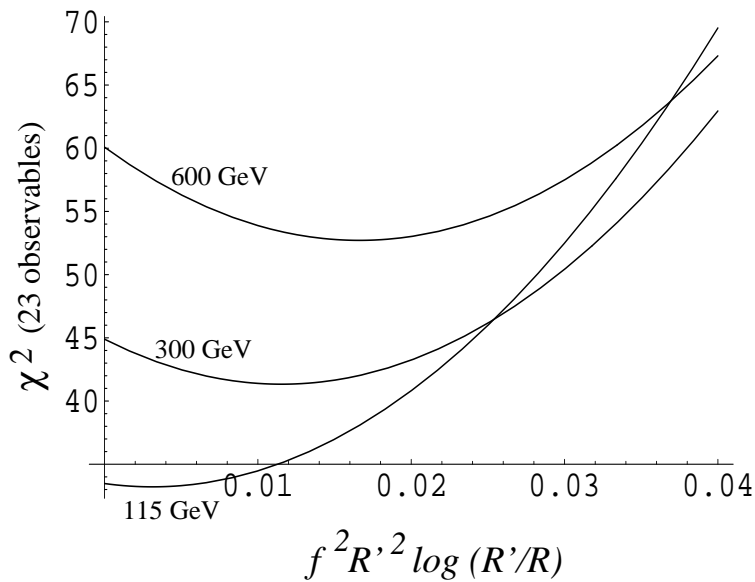}}
\caption[]{The change in $\chi^2$ in the RS model as a function 
of $R'$ and $R$, for three different values of $M_H=115, 300$,
and $600$ GeV.}
\label{fig}
\end{figure}

In the SM, the fit to data gets significantly worse when the Higgs mass is
raised. The reason is that the Higgs contributes positively to
$S$, while the data prefers a small or negative $S$. However, in our case $S$ is
negative, so one might think that a larger Higgs mass can be 
accommodated. Unfortunately 
at the same time the Higgs  also contributes
negatively to $T$, as can be seen from the approximate expressions
\cite{Sformulas}
\begin{equation}
S_{Higgs}\approx \frac{1}{12\pi} \log \left( \frac{m_H^2}{m_{H,ref}^2} 
\right), \ \ \ T_{Higgs}\approx -\frac{3}{12\pi c^2} \log \left( 
\frac{m_H^2}{m_{H,ref}^2} 
\right). \label{STHiggs}
\end{equation}
Therefore even though the agreement with $S$ can be improved by raising the
Higgs mass, $T$ will start to deviate even more. In order to see if the 
fit can be improved we have repeated it for the SM results evaluated at
$M_H=300$ GeV and $600$ GeV. In the case of the SM,
 $\chi^2$ for the 23 observables listed in
Table \ref{table} in the Appendix increases 
by about $11$ as $M_H$ increases 
from $115$ to $300$ GeV. 
If we now turn on the corrections from the RS model, 
the difference between the minimal $\chi^2$'s for the $300$ GeV and $
115$ GeV Higgs reduces to about 7.6, slightly improving the fit, 
but still outside the $95 \%$ 
confidence region ($\Delta \chi^2 = 6.2$)
of a two parameter fit (our parameters being the Higgs mass and
$X=f^2 R^{\prime\,2}\log(R'/R)$).
Hence, the Higgs mass bound is slightly relaxed but not
significantly. Assuming a $300$ GeV Higgs in turn would relax 
the limit for $R'$ to $1/R'\geq 9.0$ TeV (again assuming 
$\log R'/R =32$). For the case of the
$600$ GeV Higgs the increase in minimum $\chi^2$ is 
19, and is clearly excluded by a wide margin in 
the two parameter fit.  Assuming a $600$ GeV Higgs the bound
on  $R'$ becomes $8.2$ TeV $<1/R'< 22$ TeV.
These results are illustrated in Fig. \ref{fig}.
If one were intent on having a heavy Higgs one could add additional new
physics to the model that give positive contributions to $T$ 
\cite{PeskinWells}, but this
seems completely ad hoc in the present context.

The bound (\ref{bound1}) 
on $R'$ pushes the $W$ gauge boson resonance masses
up to around 27 TeV.  The graviton KK masses are always heavier than the 
gauge boson modes~\cite{Hooman}, for example here the lowest possible value is
around 46 TeV. Thus in the RS model with bulk gauge bosons
there will be no possible signatures from resonances in
$WW$ scattering to observe at the LHC. The focus would have to be on
the Higgs-radion sector~\cite{radion}.

To the extent that the RS model corresponds to a technicolor like
model we find these models can be consistent with experiment as long
as the resonance masses are dialed up. Unfortunately for technicolor
models there was no parameter which changed the resonance mass independently
of the electroweak breaking scale $f$.

\section{Conclusions}
We have matched the 5D RS model onto a 4D low-energy effective theory
and found large (logarithmically enhanced) 
negative contributions to $S$ and $T$.  It is
interesting to note that this is the first model to naturally
produce a large negative value for $S$~\cite{PeskinWells}; 
however this is mitigated by
the fact that the resulting bound on $R'$  from precision electroweak
measurements forces a seemingly unnatural hierarchy
of about a factor of 50 between the Higgs vev and the nominal 
electroweak scale of the RS model $1/R' \approx 11$ TeV.
This pushes the gauge boson resonances (and the graviton KK modes)
far beyond the reach of the LHC.

\section*{Acknowledgments}

We thank  T.~Bhattacharya, and J.T. thanks 
S.~Giddings, M.~Graesser, J.~Hewett, M.~Peskin, 
E.~Silverstein and S.~Thomas for useful discussions, and J.~Erler for
providing us the SM predictions used in our fit.
The research of C.C is supported in part by the NSF, and in part by the
DOE OJI grant DE-FG02-01ER41206. The research of J.E. and J.T. is supported
by the US Department of Energy under contract W-7405-ENG-36.

\section*{Appendix}

\section*{A. Predictions for  Electroweak Observables}
\renewcommand{\theequation}{A.\arabic{equation}}
In this appendix we give the predictions of a general model 
model with contributions to $S$, $T$, and $V$ for the electroweak 
precision observables. We also give in Table (\ref{table}) 
the experimental data ~\cite {LEPEWG,ErlerLang}
and the SM predictions used for our fit in Section 4.
Using the results given in~\cite{Sformulas,Burgess} as well as the
low-energy $\nu e$ couplings:
\beq
 g_{eV}(\nu e \rightarrow \nu e)=2\rho_*(s^2 - \frac{1}{4})~, \ \ \ 
 g_{eA}(\nu e \rightarrow \nu e)=-\frac{\rho_*}{2}~.
\eeq
we find the following results:
\begin{eqnarray}
 && \Gamma_Z = \left( \Gamma_Z \right)_{SM} \left(1 -3.8 \times 10^{-3} S + 0.011  T - 1.4 V\right) \nonumber \\
&&  R_e = \left( R_e \right)_{SM} \left(1 -2.9 \times 10^{-3} S + 2.0 \times 10^{-3} T-0.26 V \right) \nonumber \\
&& R_\mu = \left( R_\mu \right)_{SM} \left(1 -2.9 \times 10^{-3} S + 2.0 \times 10^{-3} T -0.26 V\right) \nonumber \\
&& R_\tau = \left( R_\tau \right)_{SM} \left(1 -2.9 \times 10^{-3} S + 2.0 \times 10^{-3} T -0.26 V\right) \nonumber \\
&& \sigma_h = \left( \sigma_h \right)_{SM} \left(1 + 2.2 \times 10^{-4} S -1.6 \times 10^{-4} T +0.021 V\right) \nonumber \\
&& R_b = \left( R_b \right)_{SM} \left(1 + 6.6 \times 10^{-4} S -4.0 \times 10^{-4} T +0.052 V \right) \nonumber \\
&& R_c = \left( R_c \right)_{SM} \left(1 -1.3 \times 10^{-3} S + 1.0 \times 10^{-3} T - 0.13 V \right) \nonumber \\
&& A_{FB}^e = \left( A_{FB}^e \right)_{SM} -6.8 \times 10^{-3} S + 4.8 \times 10^{-3} T -0.62 V\nonumber \\
&& A_{FB}^\mu = \left( A_{FB}^\mu \right)_{SM} -6.8 \times 10^{-3} S + 4.8 \times 10^{-3} T -0.62 V\nonumber \\
&& A_{FB}^\tau = \left( A_{FB}^\tau \right)_{SM} -6.8 \times 10^{-3} S + 4.8 \times 10^{-3} T -0.62 V\nonumber \\
&& A_{\tau}(P_\tau) = \left( A_{\tau}(P_\tau) \right)_{SM} -0.028  S + 0.020  T
-2.6 V \nonumber \\
&& A_{e}(P_\tau) = \left( A_{e}(P_\tau) \right)_{SM} -0.028  S + 0.020  T -2.6 V \nonumber \\ 
&& A_{FB}^b = \left( A_{FB}^b \right)_{SM} -0.020  S + 0.014  T - 1.8 V\nonumber \\
&& A_{FB}^c = \left( A_{FB}^c \right)_{SM} -0.016  S + 0.011  T -1.4 V\nonumber \\
&& A_{LR} = \left( A_{LR} \right)_{SM} -0.028  S + 0.020  T -2.6 V \nonumber \\
&& M_W = \left( M_W \right)_{SM} \left(1 -3.6 \times 10^{-3} S + 5.5 \times 10^{-3} T -0.71 V\right) \nonumber \\
&& M_W/M_Z = \left( M_W/M_Z \right)_{SM} \left(1 -3.6 \times 10^{-3} S + 5.5 \times 10^{-3} T -0.71 V\right) \nonumber \\
&& g_L^2(\nu N \rightarrow \nu X) = \left( g_L^2(\nu N \rightarrow \nu X) \right)_{SM} -2.7 \times 10^{-3} S + 6.5 \times 10^{-3} T -0.066 V\nonumber \\
&& g_R^2(\nu N \rightarrow \nu X) = \left( g_R^2(\nu N \rightarrow \nu X) \right)_{SM} + 9.3 \times 10^{-4} S -2.0 \times 10^{-4} T +0.10 V \nonumber \\
&& g_{eV}(\nu e \rightarrow \nu e) = \left( g_{eV}(\nu e \rightarrow \nu e) \right)_{SM} + 7.2 \times 10^{-3} S -5.4 \times 10^{-3} T +0.65V\nonumber \\
&& g_{eA}(\nu e \rightarrow \nu e) = \left( g_{eA}(\nu e \rightarrow \nu e) \right)_{SM} -3.9 \times 10^{-3} T -0.15 V\nonumber \\
&& Q_W(Cs) = \left( Q_W(Cs) \right)_{SM} -0.793  S -0.0090  T- 95 V 
\end{eqnarray}

\begin{table}[htbp]
\begin{center}
\begin{tabular}{|c|l|l|l|l|}\hline\hline
Quantity & Experiment & SM(115 GeV) & SM(300 GeV) & SM(600 GeV) \\\hline 
$\Gamma_Z$ & 2.4952 $\pm$ 0.0023 & 2.4965 & 2.4963 & 2.4954 \\ 
$R_e$ & 20.8040 $\pm$ 0.0500 & 20.7425 & 20.7403 & 20.7332\\ 
$R_\mu$ & 20.7850 $\pm$ 0.0330 & 20.7426 & 20.7405 & 20.7334 \\ 
$R_\tau$ & 20.7640 $\pm$ 0.0450 & 20.7879 & 20.7857 & 20.7786\\ 
$\sigma_h$ & 41.5410 $\pm$ 0.0370 & 41.4800 & 41.4774 & 41.4814 \\ 
$R_b$ & 0.2165 $\pm$ 0.00065 & 0.2157  & 0.2154 & 0.2151 \\ 
$R_c$ & 0.1719 $\pm$ 0.0031 & 0.1723 & 0.1724 & 0.1725\\ 
$A_{FB}^e$ & 0.0145 $\pm$ 0.0025 & 0.0163 & 0.0159 & 0.0157 \\ 
$A_{FB}^\mu$ & 0.0169 $\pm$ 0.0013 & 0.0163 & 0.0159 & 0.0157 \\ 
$A_{FB}^\tau$ & 0.0188 $\pm$ 0.0017 & 0.0163 & 0.0159 & 0.0157 \\ 
$A_{\tau}(P_\tau)$ & 0.1439 $\pm$ 0.0041 & 0.1475 & 0.1457 & 0.1446\\ 
$A_{e}(P_\tau)$ & 0.15138 $\pm$ 0.0022 & 0.1475 & 0.1457 & 0.1446 \\ 
$A_{FB}^b$ & 0.0990 $\pm$ 0.0017 & 0.1034 & 0.1021 & 0.1013 \\ 
$A_{FB}^c$ & 0.0685 $\pm$ 0.0034 & 0.0739 & 0.0729 & 0.0723\\ 
$A_{LR}$ & 0.1513 $\pm$ 0.0021 & 0.1475 & 0.1457 & 0.1446 \\ 
$M_W$ & 80.450 $\pm$ 0.039 & 80.3890 & 80.3775 & 80.3672\\ 
$M_W/M_Z$ & 0.8822 $\pm$ 0.0006 & 0.8816 & 0.8815 & 0.8813\\ 
$g_L^2(\nu N \rightarrow \nu X)$ & 0.3020 $\pm$ 0.0019 & 0.3039 & 0.3038 & 0.3017 \\ 
$g_R^2(\nu N \rightarrow \nu X)$ & 0.0315 $\pm$ 0.0016 & 0.0301 & 0.0301 & 0.0302\\ 
$g_{eA}(\nu e \rightarrow \nu e)$ & -0.5070 $\pm$ 0.014 & -0.5065 & -0.5065 & -0.5065\\ 
$g_{eV}(\nu e \rightarrow \nu e)$ & -0.040 $\pm$ 0.015 & -0.0397 & -0.0393 & -0.0390\\ 
$Q_W(Cs)$ & -72.65 $\pm$ 0.44 & -73.11 & -73.17 & -73.20\\ 
$m_{top}$ & 174.3 $\pm$ 5.1 & 176.3 & 185 & 192\\
\hline
\end{tabular}
\end{center}
\caption{The experimental results~\cite{ErlerLang,LEPEWG}
and the SM predictions for the various
electroweak precision observables used for the fit. The SM predictions 
are for $M_{\rm Higgs}=115,\, 300,\, 600$ GeV and $\alpha_s=0.120$ and  
calculated~\cite{Erler} using GAPP~\cite{GAPP}.}
\label{table}
\end{table}

\end{document}